\documentclass[a4paper,11pt]{article}
\pdfoutput=1 

\usepackage{jcappub} 

\usepackage[T1]{fontenc} 

\newcommand{\etal}{{et al.~}}
\newcommand{\ie}{{i.e.,~}}

\newcommand{\beq}{\begin{equation}}
\newcommand{\eeq}{\end{equation}}
\newcommand{\ber}{\begin{eqnarray}}
\newcommand{\eer}{\end{eqnarray}}
\newcommand{\lleq}{\lower0.9ex\hbox{ $\buildrel < \over \sim$} ~}
\newcommand{\ggeq}{\lower0.9ex\hbox{ $\buildrel > \over \sim$} ~}
\newcommand{\lsim}{\ \lower-1.5pt\vbox{\hhbox{\rlap{$<$}\lower5.3pt\vbox{\hhbox{$\sim$}}}}\ }
\newcommand{\gsim}{\ \lower-1.5pt\vbox{\hhbox{\rlap{$>$}\lower5.3pt\vbox{\hhbox{$\sim$}}}}\ }

\newcommand{\de}{dark energy~}
\newcommand{\cc}{{cosmological constant}}

\newcommand{\omt}{\Omega_{\rm m}^0}

\newcommand{\ome}{\Omega_{\rm {e}}}

\newcommand{\w}{w_{DE}}
\newcommand{\wdep}{w^{\prime}_{DE}}
\newcommand{\csde}{c^2_{sDE}}

\newcommand{\HH}{{\mathcal H}}
\newcommand{\HHs}{{{\mathcal H}^2}}

\newcommand{\lam}{\Lambda}

\title{\boldmath Constraining Dark Energy Perturbations: the Role of Early Dark Energy}

\author[1]{Archita Bhattacharyya,\note{Corresponding author.}}
\author{and Supratik Pal}
\affiliation{Physics and Applied Mathematics Unit, Indian Statistical Institute, 203 B. T. Road, Kolkata 700108, India}

\emailAdd{archita.bhattacharyya91@gmail.com}
\emailAdd{supratik@isical.ac.in}

\abstract{Dark Energy not only has background effects through its equation of state $\w$, but also it can cluster through its sound speed $\csde$, subject to certain conditions. As is well-known, for dynamical \de models, \de perturbations get sourced into matter perturbations through metric perturbations which is always accompanied by the term $(1+\w)$. Hence, for dynamical \de models with $\w$ close $-1$, their perturbations get almost decoupled from metric leaving nearly null imprints  on matter power spectra. Furthermore, Quintessence models with its sound speed equal to speed of light, washes out almost any inhomogeneities occurred within sub-Hubble scales, hence making detectability of \de perturbations far more difficult than already is. In this article we look for these imprints by going beyond Quintessence considering an Early Dark Energy parametrization that  not only have a non-negligible energy density at early times, but also it can achieve $\w$ far from $-1$, making \de perturbations detectable in sub-horizon scales. With the help of current datasets, we are able to constrain sound speed of \de to a low value ($\csde\sim 0.14$), along with a much higher range allowed for early \de density, with strong constraints on it ($\ome\sim 0.02$). We discuss effects of different datasets on this parametrization along with possible explanation for deviation on certain parameter(s) comparing between $\csde=1$ case and the case where it is kept open.}

\begin{document}
\maketitle
\flushbottom

\section{Introduction}
\label{sec:intro}

Although two decades have passed since it was first discovered \cite{Riess:1998cb,Perlmutter:1998np} that the expansion of our universe is accelerating at present epoch, we know surprisingly little about the exact reason behind it. Within Einstein's framework of general theory of relativity, the mysterious component \de with negative pressure is believed to be responsible for this late-time acceleration. 
Popular alternatives like cosmological constant or modified gravity are also in vogue. Over the years enormous efforts have been put in to understand its behaviour from both theoretical and observational points of view \cite{Peebles:2002gy,Sahni:2006pa,Copeland:2006wr,Uzan:2006mf,Linder:2007wa,Li:2011sd,Weinberg:2012,Bahamonde:2017ize,Huterer:2017buf}. However, in this era of cosmological precision, even the tightest constraint on equation of state \cite{Ade:2015rim} of \de gives us enough freedom to look for its origin in several avenues of physics.

Futhermore, measuring Equation of State (EoS) of \de to a great precision is certainly not the end of the story as it does not provide any explanation whether the origin of this mysterious component is realised with a canonical scalar field minimally coupled to gravity, or fields with more complicated potentials or some exotic fluid or even in a modified gravity theory. As different dynamical models of \de can have different imprints on perturbative universe, it is important to look for that signature to further discriminate among various models. At the linear level one important parameter in this regard is the speed at which pressure fluctuation of a dynamical fluid propagates \ie the sound speed of a fluid, defined by $c_s^2=\delta p/\delta\rho$. This carries much more information as it helps in describing the behaviour of perturbation; for example in case of Quintessence the fluctuation propagates at the speed of light, while for k-essence models sound speed can have arbitrary value causing interesting signatures to find out from growth of perturbations \cite{Ballesteros:2010ks,Creminelli:2009mu}.

In \cite{Sapone:2009,Sapone:2010}, the authors have described how perturbation in \de with a constant EoS affects various cosmological observations. They have analytically derived expressions to show that, while different \de models can have same EoS, they usually differ in the perturbations. The models that were discussed there, do not impact much additional to matter perturbations, as the EoS was pretty close to $-1$ for those class of models and in this limit metric perturbations decouples from \de perturbations. However, for the models where $w$ is not so close to $-1$ and especially when \de sound speed is less than speed of light, the impact can, in principle, be significantly strong.

Over the years, studies have been made to look for these observational signatures of \de through its sound speed, as well as constraining the parameter itself \cite{Bean:2003fb,Xia:2007km,Putter:2010,Calabrese:2010}, with the help of different \de models. The usual notion is that, as there is consensus about \de EoS to be pretty close to $-1$ today from latest observations, so looking for its perturbative signatures in the recent universe is not a good idea; keeping in mind that \de perturbations are always smaller than the dark matter perturbations at least by a factor of $(1+w)$. So, in order to achieve something interesting in the perturbation sector, one needs to try out particularly those \de models where its EoS can vary to have a $w$ far from $-1$
and can still be consistent with present-day observational data.

In this article we aim to search for the imprints of \de perturbations, by considering an early \de parameterization, where \de can have a non-negligible energy density even at the time of decoupling. To mention here, Early Dark Energy models (EDE, hereafter) are considerably good in this approach especially when looking for sound speed signatures on observables, as they can achieve $w$ far from $-1$ in the course of its evolution. In this article, we follow this approach, and aim for constraining \de sound speed, while searching for its non-trivial effects in the perturbative universe.

The plan of the paper is as follows: in section~\ref{sec:setup} we discuss about general \de perturbations, section~\ref{sec:model} is where we argue why Early Dark Energy models are supposed to have stronger signatures \de sound speed than the $\lam$CDM-like models, section~\ref{sec:effects} deals with the effects of non-trivial \de sound speed on observables, section~\ref{sec:res}, we present observational constraints on \de sound speed using latest datasets and finally in section~\ref{sec:con} we conclude.

\section{Dark Energy Perturbations in linear order}
\label{sec:setup}

Let us briefly discuss here the evolution of density perturbations for dark energy \cite{Sapone:2009,Sapone:2010}. In a flat FRW universe scalar perturbations in Newtonian gauge are given as,
\beq
ds^2=a^2\{-(1+2\psi)d\eta^2+(1-2\phi)dx_idx^i\} \,\,, 
\eeq
where $a$ is the scale factor, $\eta$ is conformal time, $dx^i$ is three spatial coordinates and $\phi, \psi$ are metric potentials. We consider here dark energy as an additional non-interacting fluid (along with matter and radiation), characterised by its equation of state, 
\beq
w_{DE}=\frac{p_{DE}}{\rho_{DE}} \,\,,
\eeq
and adiabatic sound speed 
\beq
c^2_{aDE} = \frac{p'_{DE}}{\rho'_{DE}}=
w_{DE}+\frac{w'_{DE}}{\rho'_{DE}/\rho_{DE}}
\eeq
The non-adiabatic contribution to sound speed can be taken care of by introducing the sound speed in its rest frame as,
\beq
c^2_{sDE} = \frac{\delta p_{DE}}{\delta \rho_{DE}} \,\,,
\eeq
where 
\beq
\delta p_{DE} = c_{aDE}^2\delta\rho_{DE} +(\csde - c_{aDE}^2)[3\HH (1+\w)\rho_{DE}]\frac{\theta_{DE}}{k^2} \,\,.
\eeq
With the help of these degrees of freedom, the perturbation equations for the dark fluid can be written as,
\ber
&& \delta^{\prime}_{DE} +3\HH (\csde-\w) \delta_{DE} + (1 + \w) \theta_{DE} +9 \HHs \left[(1 - \w)(\csde - \w) - \frac{\wdep}{3 \HH}\right]\frac{\theta_{DE}}{k^2} \nonumber\\
&& - 3(1 + \w) \phi^{\prime} =0\\
&& \theta^{\prime}_{DE}+ \HH (1 - 3\csde) \theta_{DE} - \frac{\csde}{(1+\w)} k^2 \delta_{DE} = 0 \label{eq:pert_v}\,\,.
\eer
Considering no anisotropic stress we have, $\phi=\psi$. Finally, the gravitational potential in this scenario reads as,
\beq
k^2 \phi = -4\pi Ga^2\sum_i \rho_i \left(\delta_i +\frac{3\HH (1+w_i)}{k^2}\theta_i\right) \,\,,
\eeq
where the sum runs over all the components of the universe namely, matter, radiation and dark energy.

We now discuss how \de sound speed can regulate its perturbative effects on different length scales, following more or less \cite{Putter:2010}. We begin with the scenario, where \de perturbations are within its sound horizon \ie $k >> \HH/c_{sDE}$. Now looking at Eq. (2.6) and (2.7), we see that from a $\delta''$ equation (taking derivative of (2.6) and substituting $\theta'$ from (2.7)), the small scale perturbations get washed away by its sound speed if $\csde>0$. On the other hand, in a region $\HH < k< \HH/c_{sDE}$, (assuming $\csde< 1$), \de perturbations can grow. If $\csde=1$, then \de sound horizon becomes equal to Hubble horizon, thereby making \de perturbation have observable effect on very large scales only.
We also note from Eq. (2.6), that the source term $\phi'$ is accompanied by the factor $(1+w_{DE})$. This means, for the cases where $(1+w_{DE})\rightarrow 0$, \de perturbations ($\delta_{DE}$) get decoupled from metric ones ($\phi$), and cannot anymore get sourced into matter perturbations ($\delta_m$) as depicted in Eq. (2.8). As a result, for EoS close to $-1$, even if \de has a small sound speed \ie $\csde< 1$, it does not have observational effects in matter power spectra. 

This being the case for most dynamical dark energy models, with their behavior close to $\lam$ at late times, \ie $w_0$ close to $-1$, makes it difficult to find any imprints of \de perturbations on cosmological observables via its sound speed. For instance, as shown in \cite{Putter:2010,Archidiacono:2014msa} with $w$CDM model or as in \cite{Bhattacharyya:2018fwb} with $w_z$CDM model which uses CPL parametrisation as \de model, when used to constrain perturbative behavior of dark energy, we see neither \de sound speed gets constrained by the observational data, nor it has any considerable effects on other parameters. In a  nutshell, as current observations \cite{Ade:2015rim} reveal that present day EoS for \de is more or less consistent with vacuum energy $\lam$, by allowing only a small window around $-1$, so for models of \de with $\csde=1$ and $\w\sim -1$, it is not possible to have imprints of \de perturbations if there is any, except on the largest scales, which is already dominated by cosmic variance.

What transpires from the above discussion is that, in order to detect \de perturbations on scales smaller than the horizon, one needs to consider models where $\w$ is allowed to deviate from  $-1$ to a considerable extent so that sound speed can have distinguishable influence.  In this line, EDE models turn out to be a very promising candidate, where \de can have a non-negligible energy density even at the time of decoupling. They are also known to achieve an EoS far from its present value in the past, making them ideally suitable for the purpose. This is our primary motivation of the present analysis.
In section~\ref{sec:model}, we discuss in details about the properties of this model and argue why this model seems theoretically much more appropriate in this approach.

\section{Parametrization of Early Dark Energy}
\label{sec:model}

We begin with a widely used parametrization  of early \de models. It was first introduced in \cite{Doran:2006kp}, where the fractional \de, density has been parametrized as,
\beq
\Omega_{\rm {de}}(a)= \frac{\Omega_{\rm {de}}^0 - \ome (1-a^{-3w_0})}{\Omega_{\rm {de}}^0  +\Omega_{\rm m}^0 a^{3w_0}} + \ome (1-a^{-3w_0})\,\,,
\eeq
and the corresponding equation of state is given by,
\beq
w(a) = -\frac{1}{3[1-\Omega_{\rm {de}}(a)]}\frac{d\ln \Omega_{\rm {de}}(a)}{d\ln a} + \frac{a_{eq}}{3(a+a_{eq})} \,\,,
\eeq
where $\Omega_{\rm {de}}^0$ and $\Omega_{\rm m}^0$ represent the present day fractional \de density and matter density respectively, following the relation $\Omega_{\rm {de}}^0 + \Omega_{\rm m}^0 = 1$ for a flat universe; $\Omega_{\rm e}$ is the EDE density and $a_{eq}$ is the matter-radiation equality.

\begin{figure*} 
	\centering
	$\begin{array}{cc}
	\includegraphics[width=0.45\linewidth]{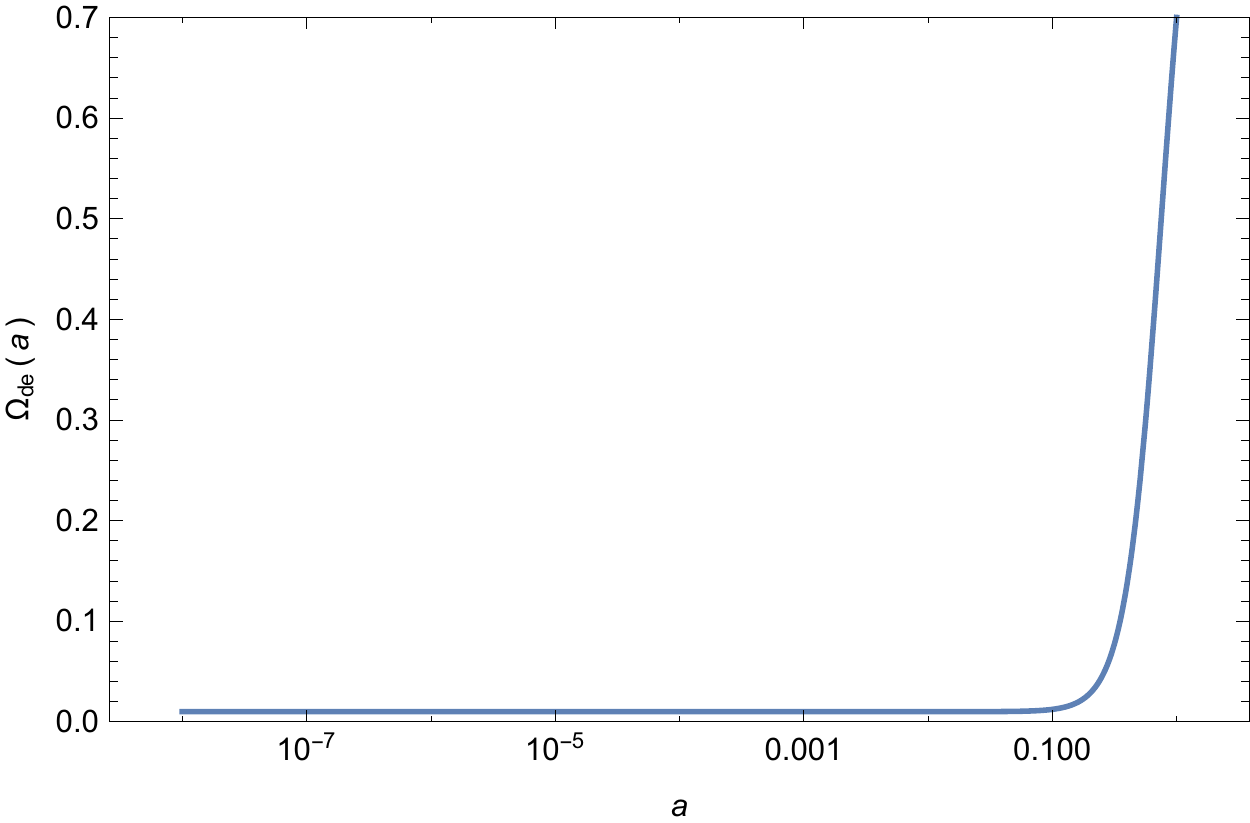} &
	\includegraphics[width=0.45\linewidth]{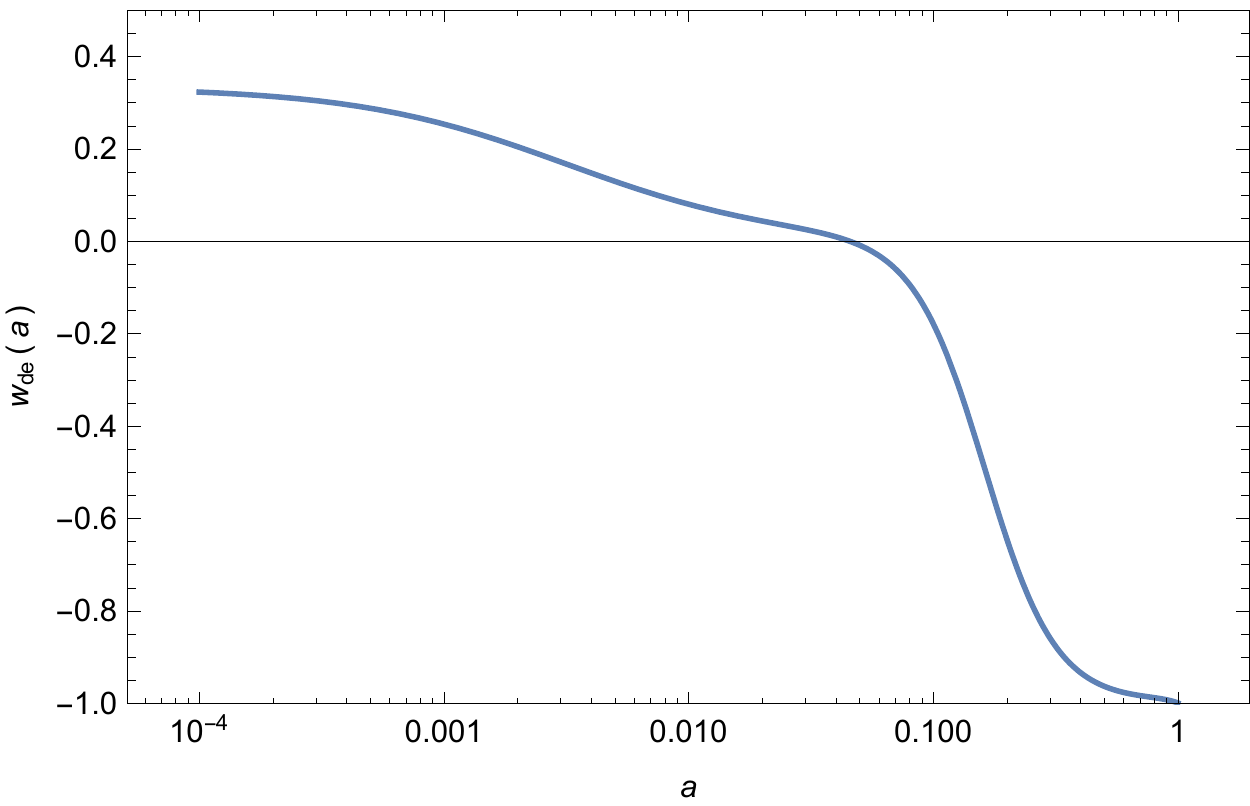} \\
	\mbox{(a)} & \mbox{(b)}
	\end{array}$
	\caption{\footnotesize
		evolution of the EDE model in its (a) energy density and (b) EoS with $\ome=0.01, w_0=-1, \Omega_m^0=0.3$ and $a_{eq}=0.0032$.}
	\label{fig:Om&wde}
\end{figure*}

In Fig.~\ref{fig:Om&wde}, we plot $\Omega_{\rm {de}}(a)$ and $w(a)$ for $w_0=-1, \ome=0.01$ and $\Omega_{\rm m}^0=0.3$. We note that \de density behaves three different ways here. In the radiation dominated era it behaves as radiation component \ie $\w \sim 1/3$, in matter dominated era as matter \ie $\w \sim 0$ and finally at present time $\w \sim w_0$, the current EoS for \de.
 This parametrization has been studied extensively in the literature \cite{Calabrese:2010,Archidiacono:2014msa,Verde:2016wmz,Lorenz:2017fgo,Calabrese:2011,Reichardt:2012,Pettorino:2013ia,Calabrese:2013lga,Bonilla:2017ygx,Caldwell:2018feo} and also by Planck collaboration \cite{Ade:2015rim}. 

Unlike the standard $\lam$CDM model, this EDE parametrization has the unique feature of having a non-negligible \de density at high-redshift, although behaving similar to $\lam$CDM at late times.  
A non-negligible \de density at early times is known to slow down the expansion rate than the standard case and hence helps ISW effect to be initiated at an earlier time. Furthermore, as is seen from fig.~\ref{fig:Om&wde}(b), this \de model is allowed to have $\w$ far from $-1$ in the course of its evolution, which is a necessary condition to obtain a clearer signature of \de sound speed, that we already discussed in section~\ref{sec:setup}. 

The Hubble parameter here reads as,
\beq
\frac{H^2 (a)}{H_0^2} = \frac{\Omega_{\rm m}^0 a^{-3}+\Omega_{\rm {rel}}^0a^{-4}}{1-\Omega_{\rm {de}}(a)}\,\,,
\eeq
where $\Omega_{\rm {rel}}^0$ is fractional energy density of relativistic particles at present day.

It is worth mentioning here that, the parameterization described above is related to the model-independent parameterization of {\it Quintessence} scalar field models given by \cite{Corasaniti:2002vg},
\beq
w_Q(a)= w_Q^0 + (w_Q^m - w_Q^0) \times \frac{1+e^{\frac{a_c^m}{\Delta_m}}}{1+e^{-\frac{a-a_c^m}{\Delta_m}}} \times \frac{1-e^{-\frac{a-1}{\Delta_m}}}{1-e^{\frac{1}{\Delta_m}}}\,\,,
\eeq
where $w_0^Q \equiv w_0$, $w_Q^m$ is the \de EoS at early times, which is always zero for the parameterization we have chosen here, $a_c^m$ is the scale factor at transition from $w_Q^m$ to $w_Q^0$ and $\Delta_m$ is the width of transition. See ref \cite{Doran:2006kp} for more details on this. 
For models with $w_Q^m$ close to $w_0$, their behavior are close to $\lam$ during its evolution, and hence making its perturbative effects unobservable as discussed in section \ref{sec:setup}. Here, $w_m^Q$ is zero all the time, which makes the difference with $w_0$ the most, thereby making this model potent for seeing its effects on perturbative universe.

Below we describe theoretically how this parametrization in eq. (3.1) suits in our approach of constraining \de sound speed through its enhanced effect of \de perturbation on observables.

\section{Effect of Dark Energy Perturbations on Observables}
\label{sec:effects}

We now discuss, how CMB and matter-power spectra get affected by \de sound speed. In CMB spectra, effects of \de enters mainly through ISW effect \cite{Weller:2003hw}, which is basically decaying of gravitational potentials starting at the onset of \de domination. Keeping in mind the main contribution to ISW comes from the difference in cosmological expansion, we want to discriminate effects of \de sound speed, aiming to constrain it from latest observations. In order to realise the effect of sound speed on 
CMB and matter-power spectra, we investigate two extreme cases as demonstrative examples, namely, $\csde=1$ and $\csde=0$, and compare them with the vanilla $\lam$CDM model.

\begin{figure*} 
	\centering
	$\begin{array}{cc}
	\includegraphics[width=0.5\linewidth]{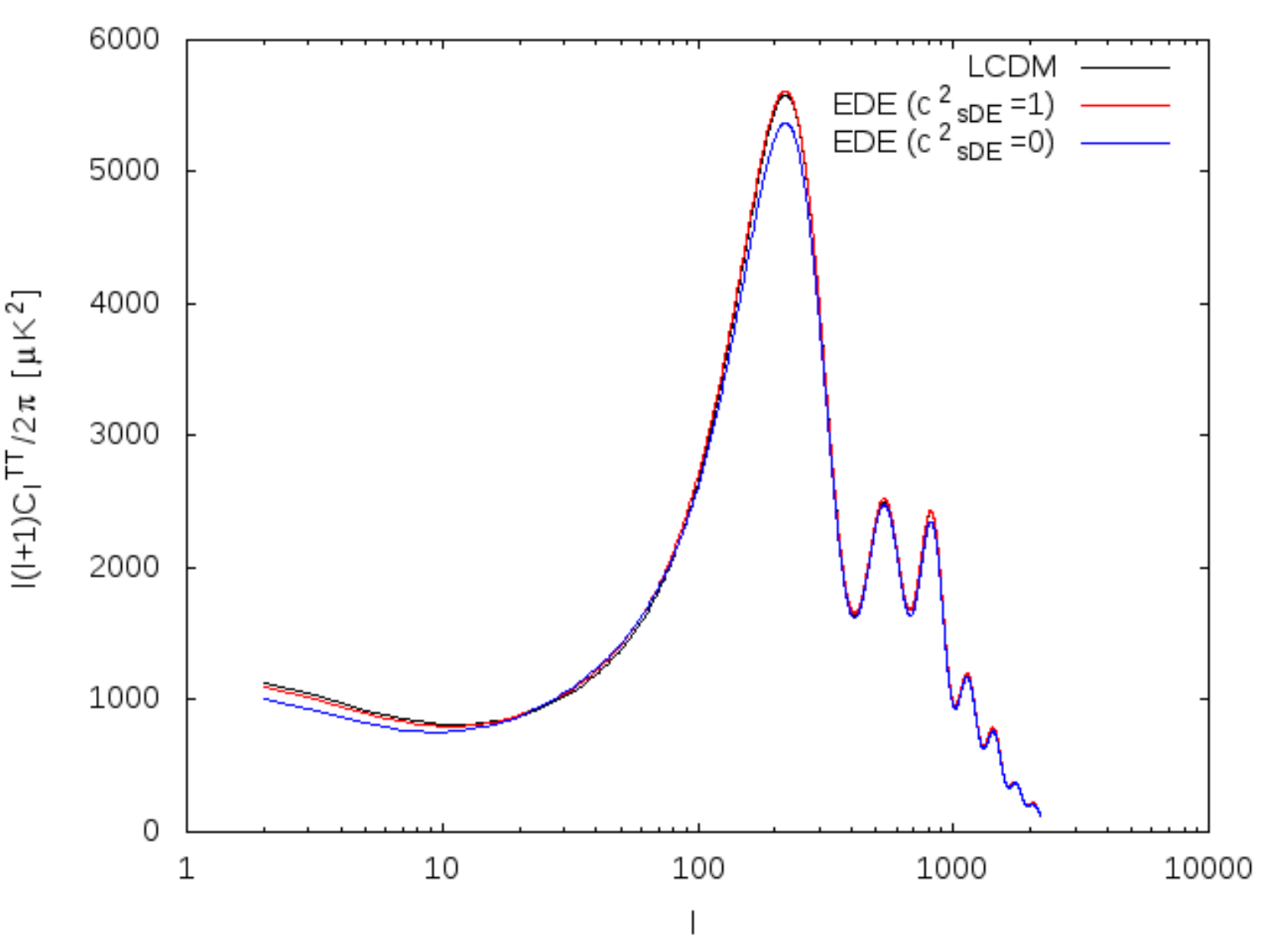} &
	\includegraphics[width=0.5\linewidth]{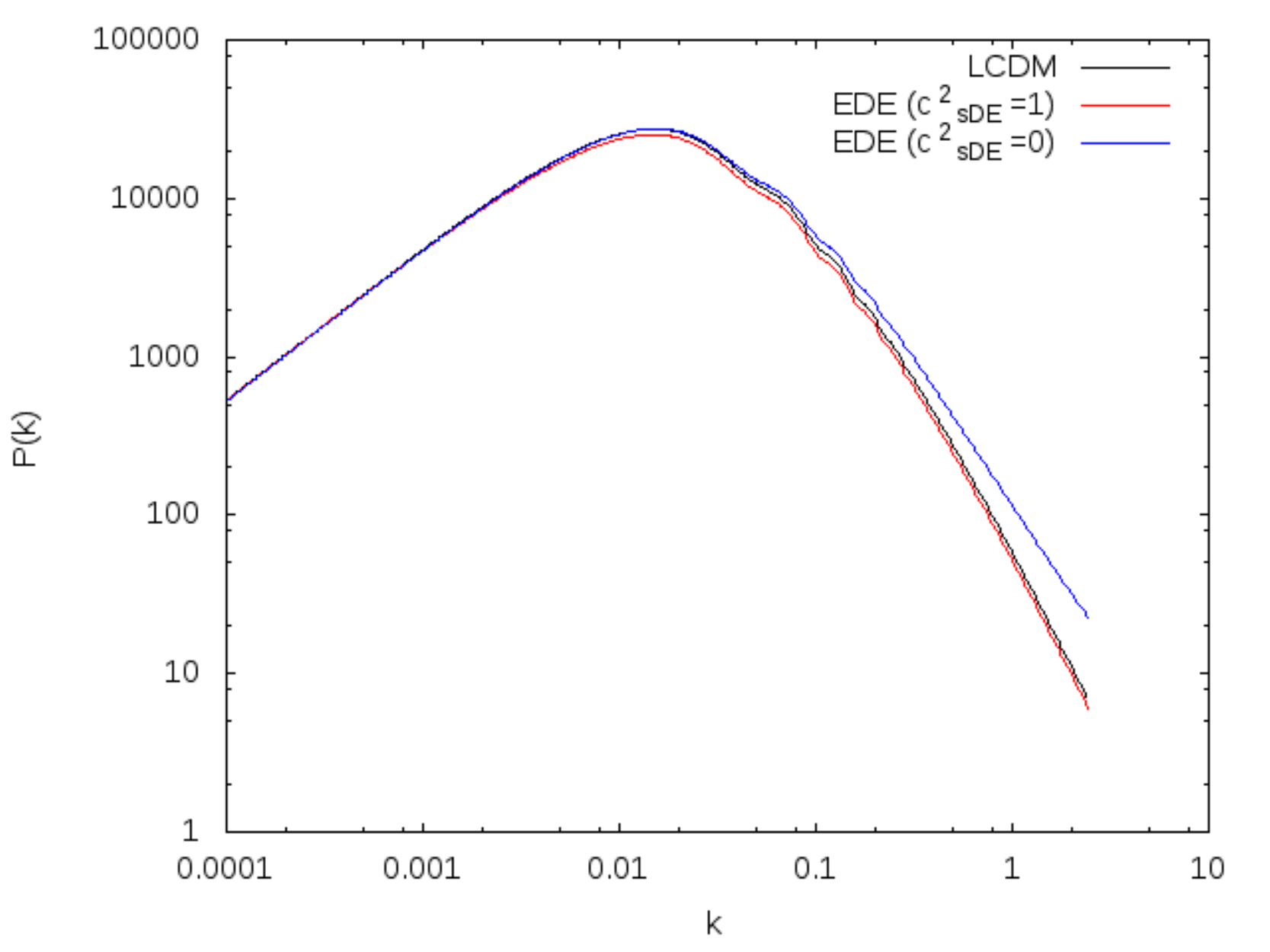} \\
	\mbox{(a)} & \mbox{(b)}	
	\end{array}$
	\caption{\footnotesize
		CMB and matter-power spectra EDE model with $\csde=1$ (red lines) and with $\csde=0$ (blue lines). The LCDM curves (black lines) are given for comparison.}
	\label{fig:Cl&Pk}
\end{figure*}

Fig. \ref{fig:Cl&Pk}(a) shows the effect of varying sound speed for EDE model on CMB TT spectrum. The black lines depict the $\lam$CDM curve, for which ISW power is strongest. For EDE model with sound speed equal to speed of light \ie $\csde=1$, \de cannot cluster due to its large sound speed. As a result ISW power remains near $\lam$CDM in this case (red lines), the only difference caused by a different expansion rate than $\lam$CDM. However, for the EDE case with $\csde=0$,  \de clusters significantly, contributing to the gravitational potential $\phi$, thereby somewhat compensating for the decay of the potential for the background effect (blue lines). Hence, ISW power stays the smallest in this case. 

Apart from the effects in the largest scales where cosmic variance dominates the signal immensely, in the case of EDE scenario, there are non-negligible amount of \de present at early times, causing decay of gravitational potential to be initiated at earlier times than usual, resulting in signatures detectable at larger-$\ell$s, where the error bars are sufficiently small. In addition to this, presence of \de density at early times can change the amplitude of acoustic oscillations before recombination, as seen from fig. \ref{fig:Cl&Pk}(a). If the sound speed is increased, the perturbations in EDE are strongly suppressed, causing larger amplitude of acoustic oscillations.

The matter power spectrum is affected by the \de sound speed through potential perturbations induced by \de clustering as shown in eq. (2.8). We keep in mind here that, \de perturbations ($\delta_{DE}$) itself remain smaller by matter perturbation ($\delta_m$) by a factor $(1+\w)$ in sub-horizon scales, where the \de perturbations stop growing if $\csde>0$. On the other hand, in super-horizon scales $\delta_{DE}$ is further suppressed by the factor $1/(1-3\w)$. These factors suppress \de perturbations considerably when $\w\sim -1$. For the models where $\w$ is far from $-1$, however, one expects significant clustering \de and hence becoming more promising in constraining \de sound speed over other dynamical models. 
Fig. \ref{fig:Cl&Pk}(b) shows the effect of changing sound speed on matter power spectrum for EDE model. We see that the curve for $\lam$CDM and that for EDE with $\csde=1$ have a little difference between them. This is expected, as \de imhomogeneities get washed away because of its large sound speed, making its contributions almost null in matter-power spectrum. On the other hand, in $\csde=0$ case, \de clusters significantly, thus enhancing power as seen from the blue lines.

With these results in mind, we expect to see clearer signatures of \de perturbations when considering this parametrization. In the next section, we shall do a more realistic analysis keeping the sound speed of EDE as a free parameter that will reflect its effects on different observations. We will therefrom see if we can
constrain this sound speed and how that is going to affect the other cosmological parameters with the help of latest datasets.

\section{Method, Datasets, Results and Discussions}
\label{sec:res}

Having convinced ourselves of the prospects of the parametrization under consideration,
we will now proceed to constrain \de sound speed along with other cosmological parameters from different observations. To obtain this we shall perform a Markov Chain Monte Carlo likelihood analysis with the help of publicly available package CosmoMC, with a modified version of CAMB. As our datasets, we use:
\begin{itemize}
	\item CMB: Planck TT and low-$\ell$ data from the Planck 2015 data release \cite{Ade:2015rim}.
	\item Galaxy BAO: Measurements from 6dFGS at $z=0.106$ and MGS at $z=0.15$
	from SDSS, as well as the CMASS and LOWz samples from BOSS DR12 at
	$z=0.38,0.51$ and $0.61$ \cite{Alam:2016hwk}.
	\item SN Ia: SN Ia data from Pantheon sample, which consists of 279 PS1 (Pan-STARRS1) SN Ia ($0.03<z<0.68$), with useful distance estimates of SN Ia from SDSS, SNLS, various low-z and HST samples. The combined dataset comprises of 1048 SN Ia ranging from $0.01< z <2.3$ \cite{Scolnic:2017caz}.
	\item DES Y1: First year Weak Lensing data from Dark Energy Survey \cite{Abbott:2017wau}.
\end{itemize}

In order to investigate for the effects of different datasets on $\csde$  and, in turn, on the other parameters, we first constrain the EDE model by fixing $\csde$ to unity, that reflects the behavior of canonical scalar field models. Later on we keep it as a free parameter in our code and examine the possible deviation of the results from the $\csde =1$ scenario as well as possible constraints on the parameter itself, and on other cosmological parameters. 
 
In table \ref{tab:ede} we show the results for the two cases where $\csde$ is kept fixed at unity and an open parameter simultaneously, so that the results are easier to compare. The parameters of interest here are:
$w_0,\ome,\csde,\omt,H_0,\sigma_8$. We see that using Planck data alone has constrained $\ome < 0.0036$ at 95\% confidence level, when $\csde$ is fixed at 1, however keeping the constraint on $w_0$ too broad, at $<-0.45$. This also shows preference for a higher $\omt$ and lower $H_0$, as seen from Table \ref{tab:ede}, with its $\sigma_8$ value consistent with $\lam$CDM within 1$\sigma$.

Allowing $\csde$ as an open parameter we first see that, $\ome$ is now strongly constrained at $\gtrsim 0.018$ at 1$\sigma$, without much change in $w_0$. Along with this, $\csde$ is now constrained at a much lower value of $\sim 0.14$, than its canonical scalar field value. Consequently, we do not see much change in $\omt$ and $H_0$ value, but a slightly lower value of $\sigma_8\sim 0.73$ is allowed. The reason behind this is as follows: as we have seen in fig. \ref{fig:Cl&Pk}(b), an EDE model with $\csde=1$ suppresses the \de perturbation mostly, hence we expect less clustering than in $\lam$CDM, which would reflect in a lower $\sigma_8$ value. On the contrary, in the case for $\csde=0$, we expect \de to cluster itself, hence an increased power as seen from fig. \ref{fig:Cl&Pk}(b). But here we notice, with a low $\csde$ chosen by the observation, value of $\ome$ is now increased to a much higher value of $\sim 0.02$. This in turn, affects $\sigma_8$, with its lower value afforded mostly by enhanced background effect, with much larger $\ome$ than the previous case. As a result, we get a lower $\sigma_8$ even with a small sound speed of dark energy. This definitely puts new limit on early \de density alongside low $\csde$ than earlier analysis.

This result is also important from the viewpoint of constraining \de perturbation. We remind the reader here that, it has not been possible to constrain dark energy sound speed before. The reason behind this directly follows from our argument in section \ref{sec:setup}. As for most dynamical dark energy models, their behavior resembles $\lam$, with an EoS close to $-1$ throughout cosmological evolution, their perturbative behavior ($\csde$) are very poorly constrained. By choosing this EDE model, we are allowing $\w$ to deviate from $w_0$ fairly and hence expect observation to reveal stronger effect than before. From our analysis we see that Planck CMB data has constrained \de sound speed at $\sim 0.14$, which is much lower than those for canonical scalar field \de models.

\begin{table*}
	\centering
	\begin{tabular}{cccccccc}
		\hline \hline
		Data&Model&$w_0$&$\ome$&$\csde$&$\omt$&$H_0$&$\sigma_8$\\
		\hline
		Planck&EDE&$<-0.45$&$<0.0036$&1&$0.45_{-0.08}^{+0.04}$&$57.6_{-2.8}^{+5.1}$
		&$0.762_{-0.035}^{+0.064}$\\
		&$+\csde$&$<-0.50$&$0.0211_{-0.003}^{+0.003}$&$0.14_{-0.006}^{+0.006}$
		&$0.47_{-0.07}^{+0.03}$&$57.3_{-2.3}^{+4.0}$&$0.732_{-0.03}^{+0.05}$\\
		\hline
		Planck&EDE&$<-0.93$&$<0.0028$&1&$0.33_{-0.009}^{+0.009}$&$65.5_{-0.67}^{+0.75}$
		&$0.798_{-0.018}^{+0.018}$\\
		+BAO&$+\csde$&$<-0.94$&$0.018_{-0.004}^{+0.004}$&$0.135_{-0.007}^{+0.008}$
		&$0.33_{-0.009}^{+0.008}$&$65.6_{-0.6}^{+0.7}$&$0.757_{-0.018}^{+0.018}$\\
		\hline
		Planck&EDE&$<-0.96$&$<0.0032$&1&$0.35_{-0.01}^{+0.01}$&$64.2_{-0.8}^{+0.8}$
		&$0.810_{-0.018}^{+0.018}$\\
		+SN Ia&$+\csde$&$<-0.97$&$0.019_{-0.003}^{+0.003}$&$0.137_{-0.007}^{+0.007}$
		&$0.36_{-0.013}^{+0.013}$&$63.8_{-0.8}^{+0.8}$&$0.77_{-0.017}^{+0.017}$\\
		\hline
		Planck&EDE&$<-0.96$&$<0.0026$&1&$0.32_{-0.007}^{+0.007}$&$65.9_{-0.6}^{+0.6}$
		&$0.800_{-0.018}^{+0.018}$\\
		+BAO+SN Ia&$+\csde$&$<-0.97$&$0.018_{-0.003}^{+0.003}$&$0.135_{-0.007}^{+0.008}$
		&$0.32_{-0.007}^{+0.007}$&$66.0_{-0.5}^{+0.6}$&$0.758_{-0.019}^{+0.017}$\\
		\hline
		Planck&EDE&$<-0.82$&$<0.0043$&1&$0.33_{-0.02}^{+0.01}$&$64.6_{-1.0}^{+2.0}$
		&$0.764_{-0.016}^{+0.023}$\\
		+DES&$+\csde$&$<-0.83$&$0.021_{-0.003}^{+0.003}$&$0.137_{-0.007}^{+0.007}$
		&$0.35_{-0.02}^{+0.01}$&$63.6_{-0.9}^{+1.8}$&$0.733_{-0.015}^{+0.02}$\\
		\hline
	\end{tabular}
	\caption{\footnotesize
		Mean with $1\sigma$ errors and 95\% credible regions for upper limits on different cosmological parameters for different combination of datasets. For each (set of) observations, the first row depicts the results when $\csde$ is fixed at 1, and the second row those when $\csde$ is kept as an open parameter.}
	\label{tab:ede}
\end{table*}

We shall now add other datasets to our analysis, and find out its possible effects on the parameters under consideration. Adding BAO data to the EDE scenario, has helped shrink $w_0$ closer to its $\lam$-value \ie $<0.93$ at 95\% CL (Table \ref{tab:ede}),  and made the region for $\ome$ slightly narrower $<0.0028$. Also, BAO data brings $\omt \sim 0.33$, thereby breaking the degeneracy between $H_0$ and $\omt$ as in CMB data alone, and hence improving constraint on $H_0$ closer to its $\lam$CDM value. We also notice, $\sigma_8$ has now been pushed to a higher value of $\sim0.79$, not much different the standard case. For the EDE+$\csde$ case however, we again notice that, without changing $w_0$, $\ome$ is now constrained strongly, with $\ome\gtrsim 0.014$ at 1$\sigma$, with $\csde\sim 0.135$. In this case also, $\omt$ and $H_0$ remain unaltered when compared to $\csde=1$ scenario, with a lower allowed region for $\sigma_8$ only.

With the addition of SN Ia data, $w_0$ is narrowed down closer to $-1$ as expected, while keeping $\ome<0.0032$, slightly higher than the previous case. With these, the constraints on $\omt$ and $H_0$ are little different than what we obtained for Planck+BAO case, with slightly higher $\omt$ and lower $H_0$. While $\sigma_8$ value remains close to the standard $\lam$CDM value. Adding $\csde$ again constrains $\ome\gtrsim 0.016$ at 1$\sigma$ and $\csde\sim 0.137$. As in the previous cases, this has only moved the parameter $\sigma_8$ to lower value of $\sim 0.77$ without altering other parameters much.

We also add the Weak Lensing data from DES Y1 for the first time to constrain this model, to see that, the EDE constraints are little broader than the other cases, such as, $w_0<-0.82$ and $\ome<0.0043$ in EDE case. Adding $\csde$ to vary as well, we notice this datasets combination also behave similarly as the previous cases. Constraint on $\ome$ has now become stronger with $\gtrsim 0.021$ and a small sound speed $\sim 0.137$. While DES already constrain $\sigma_8$ a little lower than Planck (see ref. \cite{Handley:2019wlz}), this parametrization shows further diminishing effect on this parameter.

\begin{figure*}
	\includegraphics[width=0.9\linewidth]{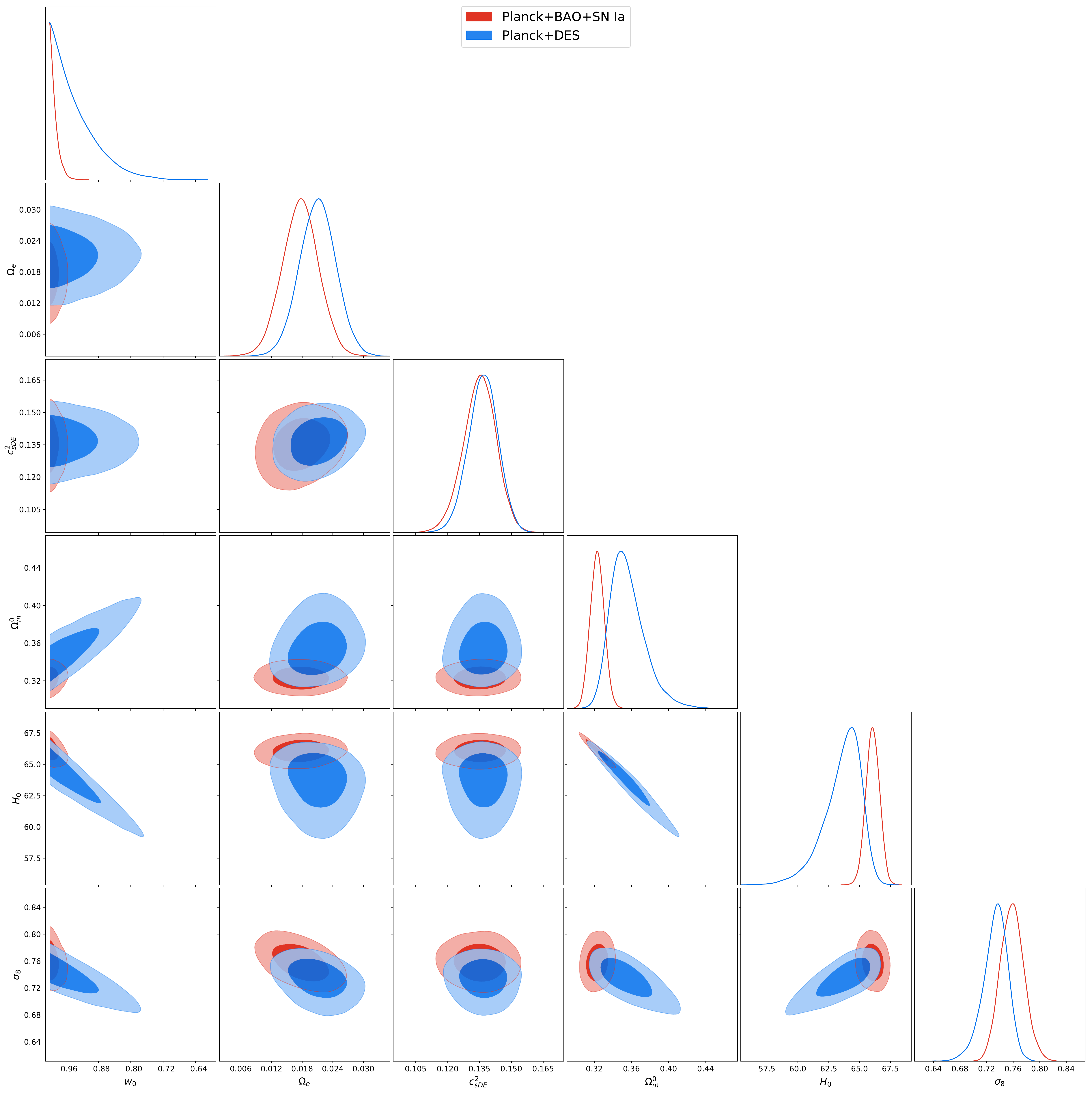} \\
	\caption{\footnotesize
		1,2$\sigma$ confidence regions and likelihoods on $w_0,\ome,\csde,\omt,H_0,\sigma_8$ using Planck+BAO+SN Ia for (red) and Planck+DES (blue).}
	\label{fig:ede_1D&2D}
\end{figure*}

The strongest constraint on $\ome$ comes when Planck, BAO and SN Ia data are combined together as seen from Table \ref{tab:ede} at $\ome\sim 0.0026$ in the EDE case, with the other parameters not much different from standard case. With the addition of $\csde$ much larger value is allowed, $\ome\gtrsim 0.015$, with Planck+BAO+SN Ia data, along with $\csde\sim 0.135$. It is intriguing to notice that, all the datasets used above, constrain the sound speed of \de when considering this parametrization, to a much lower value than its canonical scalar field value, with a higher early \de density, without affecting present day EoS for dark energy. We also see noticeable effect on the parameter $\sigma_8$, reflecting enhanced backgound effect of \de (with increased $\ome$) rather than its clustering effect on this parameter. This change in $\sigma_8$ could be a discriminatory signature of EDE models, when analysing matter-power spectra from different observations. We keep in mind here that, Planck+BAO+SN Ia data are not able to achieve $\sigma_8$ as low as this when using other \de models (such as CPL parameterization), as pointed out in \cite{Bhattacharyya:2018fwb}. So, EDE models with a low sound speed might come in handy when dealing with the $\sigma_8$ tension between CMB and Weak Lensing or cluster counts measurements. We show the correlations between the EDE parameters and on $\omt,H_0,\sigma_8$ as well as 1D likelihood plots in Fig. \ref{fig:ede_1D&2D}.

To summarise our findings from the above analysis are as below:
\begin{itemize}
	\item A small amount of \de density at early times is allowed by all the datasets used here, consistent with previous analyses. When considering EDE with $\csde=1$, early \de density has a limit of $< 0.0026$ using Planck+BAO+SN Ia.
	\item Adding DES Y1 data with Planck, allows a bigger parameter space for EDE parameters, such as, $w_0<-0.82$ and $\ome<0.0043$.
	\item Allowing $\csde$ to vary, has helped mean $\ome$ to move to a much higher value of $\sim 0.02$ and putting a strong constraint in it rather than an upper limit, with all the datasets used above; along with  dark energy sound speed now constrained to a much lower value of $\sim 0.14$.
	\item We also find that a lower $\sigma_8$ is now allowed, than achieved by Planck+BAO+SN Ia with other dynamical \de models. This might help with the $\sigma_8$ tension there is, between Planck CMB data and Weak Lensing measurements within standard $\lam$CDM model.
	\item What is revealed from the above analysis is that, it is important to look for non-trivial role of \de when considering dynamical models, as they are able to bring out interesting features and setting it to 1 a priori may not be a good idea. 
\end{itemize}

As a bottomline we state that, the simple $\lam$CDM model, commensurate with current observations, might be a mere consequence of restricting the parameter space tightly, whereas, allowing more degrees of freedom may help understanding deeper underlying physics. For instance, as we found out in this analysis, an EDE model with a constant sound speed, seems perfectly consistent with the current data, for which it is possible to constrain its sound speed strongly to a low value. This model is certainly capable of addressing richer physics than canonical scalar field models of \de characterised by $\csde=1$, and needs further investigation with future datasets.

\section{Conclusion}
\label{sec:con}

Dark Energy dominates the present universe comprising nearly two-third of its energy densities. Although, current data are in favor of \cc ($\lam$) being the most suitable \de candidate, enough room is left for several dynamical models to be consistent as well. At this stage, when new datasets are being released with many more to come, it is important to extract out as much information as they are capable of providing. In this line, when confronting dynamical \de models, one should not only concentrate on constraing EoS of \de to a more precise value, but also investigate for richer microphysics such as its sound speed, to determine its perturbative behavior.

As has been shown by the present authors as well as others before, current datasets are not able to put any constraints on \de sound speed for near $\lam$CDM cases. This is mainly because for $\w\sim -1$, \de perturbations decouple from metric ones, leaving no signatures of its perturbations, if any. Moreover, quitessence \de models characterised by $\csde=1$, does not allow for much imprints except on the largest scales, already dominated by cosmic variance. In this work, we attempted searching for imprints of perturbations in \de through its sound speed and its effects for EDE. The EDE parametrization we chose has non-negligible energy density in the early universe, as well as it can achieve EoS which is further from $-1$ than other dynamical models. Moreover, we allowed its sound speed to vary freely (though constant), to look for beyond quintessence degrees of freedom.

Using various current observations, such as Planck CMB, BAO, SN Ia and DES Y1, we found it is possible to contrain \de sound speed with the help this EDE parametrization. Nevertheless, we have found all the combinations of datasets we have used here, appear to prefer a low sound speed of \de. We also noted, opening up of the parameter space, has helped $\ome$ to move to a greater value which is almost 1$\sigma$ away from its allowed region in the case where $\csde$ was fixed at 1. As a consequence of freely varying sound sound speed, lower $\sigma_8$ regions are allowed, which could be distinct feature of these type of models, in clustering observations. These results might mean, the explanation of late-time acceleration, indeed lies in a richer physics with even beyond quintessence degrees of freedom. The simplified $\lam$-description of \de, which is spatially and temporally constant, might be the result of overly constraning the parameter space. These areas need much exploration with more future data to pin down on our understanding of dark universe.

\acknowledgments

We gratefully acknowledge use of the publicly available
code CosmoMC. We also thank computational facilities of ISI
Kolkata. AB thanks DST, India for financial support through
INSPIRE fellowship DST-INSPIRE/IF150497.

\end{document}